\documentclass[]{jfm_notyet}

\usepackage{natbib}
\usepackage{bm}
\usepackage{amssymb}
\usepackage{graphicx}
\usepackage{color}
\usepackage{amsmath}
\usepackage{amsfonts}
\usepackage{hyperref}
\usepackage{bm}
\usepackage{latexsym}
\usepackage{epsfig}

\newcommand{\be}{\begin{eqnarray}}
\newcommand{\en}{\end{eqnarray}}

\newcommand{\ud}{\mathrm{d}}

\newcommand{\etal}{{\em et~al.\/}}
\renewcommand{\vec}[1]{{\bm #1}}

\title{\emph{Sticky} elastic collisions}
\author[J.\ Bec, S.\ Musacchio, and S.\ S.\ Ray]
{J\ls \'E\ls R\ls \'E\ls M\ls I\ls E\ns B\ls E\ls C$^1$,\ns S\ls T\ls
  E\ls F\ls A\ls N\ls O\ns M\ls U\ls S\ls A\ls C\ls C\ls H\ls I\ls
  O$^2$,\ns \newline\and S\ls A\ls M\ls R\ls I\ls D\ls D\ls H\ls I\ns
  S\ls A\ls N\ls K\ls A\ls R\ns R\ls A\ls Y$^{1,3}$} 
\affiliation{
$^1$
Laboratoire J.L.\ Lagrange UMR\,7293, 
Universit\'e de Nice Sophia Antipolis, 
CNRS, Observatoire de la C\^ote d'Azur, 
BP4229, 06304 Nice Cedex 4, France.
\\[\affilskip]
$^2$
Laboratoire J.A.\ Dieudonn\'e UMR\,7351, 
Universit\'e de Nice Sophia Antipolis, CNRS,
Parc Valrose, 06108 Nice, France.
\\[\affilskip]
$^3$ International Centre for Theoretical Sciences, Tata Institute of
Fundamental Research, Bangalore 560012, India.}

\begin{document}

\maketitle

\begin{abstract}
  The effects of purely elastic collisions on the dynamics of heavy
  inertial particles is investigated in a three-dimensional random
  incompressible flow.  It is shown that the statistical properties of
  inter-particle separations and relative velocities are strongly
  influenced by the occurrence of \emph{sticky elastic
    collisions\/}\,---\,particle pairs undergo a large number of
  collisions against each other during a small time interval over
  which, hence, they remain close to each other.  A theoretical
  framework is provided for describing and quantifying this phenomenon
  and it is substantiated by numerical simulations.  Furthermore, the
  impact of hydrodynamic interactions is discussed for such a system
  of colliding particles.
\end{abstract}
\maketitle

\section{Introduction}
\noindent Dust, droplets, bubbles, and other finite-size particles
suspended in turbulent flows are common in nature \citep[see,
e.g.,][]{Csanady,Abraham,Shaw}.  Their statistical properties are very
different from that of tracers, i.e.\ point-like particles with the
same mass density as the advecting fluid. Indeed, when the suspended
particles have a finite size and a density different from that of the
fluid, inertial effects become important. Consequently, the motion of
particles starts differing from the underlying flow.  This results in
intricate correlations between the particle positions and the geometry
of the turbulent flow. Heavier particles are expelled from vortical
structures while lighter particles concentrate in their cores. A
consequence of these mechanisms is the presence of strong fluctuations
in the spatial distribution of particles. This phenomenon, known as
\emph{preferential concentration}, has been the subject of extensive
research in fluid dynamics for the last decades
\citep[see][]{Douady,SE91}.

Another consequence of inertia is that particles are likely to be very
close to each other with large velocity differences. The process
leading to such events, known either as the sling effect
\citep[][]{falko_nature} or the formation of caustics
\citep[][]{Wilkinson2005}, has been extensively measured and studied
during the past ten years \citep[see,
e.g.,][]{Bec2010,Salazar2012}. The presence of such spatial
inhomogeneities is known to strongly alter possible interactions
between particles. Since the pioneering work of \cite{Saffman}
motivated by coalescences of cloud droplets, much work has been
devoted to understanding the rate at which heavy inertial particles
collide. A commonly adopted approach consists in assuming that the
inter-collision time is much longer than the convergence timescale of
particle dynamics to a statistically stationary regime. This premise,
which is asymptotically true in the limit of very dilute suspensions,
entitles counting collisions without having to effectively perform
them. The frequency at which such \emph{ghost particles} collide is
then a time-independent statistical observables that can be quantified
as a function of particle sizes and response times.  Several studies
have assumed this ghost-particle hypothesis in order to estimate
collision rates of heavy inertial particles in turbulent flows
\citep[see, e.g.,][]{sundaram97,falko_nature}. However, little is
known about the limits of such an approach. For instance, an important
statistical weight is given to events when two concentrated clouds of
particles cross each other with a large velocity difference. It is
clear that the very-dilute approximation should then fail and that
multiple-collisions are likely to occur. We report in this paper
results on the effects of actual collisions on the dynamics and
statistics of inertial particles transported by a non-stationary fluid
flow.

The simplest framework for treating short-range collective effects in
the particle dynamics is to consider an ensemble of hard spheres that
are suspended in a prescribed flow and which undergo purely elastic
(momentum and energy preserving) collisions with each other.  Such a
system, in the absence of any underlying fluid transport, has long
been a paradigmatic model in statistical mechanics.  It is the basis
of the kinetic theory of gasses, in which it is assumed that the
kinetic energy is conserved both by the collisions and by the dynamics
of each particle.  If we allow for a certain amount of inelasticity in
the collisions a suitable model for granular gasses is
obtained~\citep[see][for a review]{goldhirsch2003rapid}.  A peculiar
behaviour of such granular systems is the spontaneous aggregation of
particles.  Since inelasticity implies the loss of a finite percentage
of kinetic energy at each collision, the particles can eventually
stick together leading to the formation of large clusters. However,
this effect occurs also in settings that are not common in studies of
granular media.  Indeed, as we will see in this work, it suffices that
kinetic energy dissipation occurs not via collisions but rather
through individual particle dynamics. This is the case for heavy
inertial particles whose motion is dominated by viscous damping and
which undergo purely elastic collisions.  We show that the clustering
phenomenon emerging in such systems originates from what we call {\it
  sticky elastic collisions}. During these events, the particles
bounce many times against each other and energy is dissipated during
their motion between successive collisions.  This mechanism has strong
influences on the statistics of inter-particle distances and relative
velocities.  Furthermore, to validate the presence of this effect in
real settings, we study the influence of hydrodynamic interactions and
show that they cannot prevent sticky elastic collisions from occurring.

This paper is organised as follows. In \S\ref{sec:model}, we discuss
the equations of motion of the inertial particles and recall some key
results, in the absence of any collisions, on preferential
concentration.  In \S\ref{sec:sticky}, we present numerical results
and provide an asymptotic analysis for the phenomenon of sticky
elastic collisions in the absence of hydrodynamic interactions. We
then discuss the effect of hydrodynamic interactions in
\S\ref{sec:hydro} by considering far--field interactions.  We make
some concluding remarks and summarise our results, as well as provide
a perspective for future work, in \S\ref{sec:concl}.

\section{The Model}
\label{sec:model}
To set the stage, we begin by recalling the basic physics of a system
of $N$ small hard spheres which are in a random, time-dependent,
incompressible fluid field $\vec{u}(\vec{x},t)$ and are subject to
viscous dissipation.  When such particles have a small Reynolds number
and are much heavier than the fluid, they interact with the flow by
Stokes viscous drag and their trajectories $\vec{x}_i(t)$ are
determined by Newton's law:
\begin{equation} 
\dot{\vec{x}}_i = \vec{v}_i,\quad \dot{\vec{v}}_i =
-\frac{1}{\tau}\left[\vec{v}_i - \vec{u}(\vec{x}_i,t) \right]\;\;\; i
\in [1,N]
\label{eq:1}
\end{equation}
where $\tau$ is the viscous-drag relaxation (Stokes) time defined via
$\tau =2 \rho_p a^2 /(9 \rho_f \nu)$, where $\rho_p$ is the particle
density, $\rho_f$ the fluid density, $\nu$ its kinematic viscosity,
and $a$ is the particle radius. The Stokes number is a measure of the
inertia of the particle and is defined as $St=\tau/\tau_\mathrm{f}$,
where $\tau_\mathrm{f}$ is a characteristic timescale of the fluid
flow.  In this work, we additionally introduce interactions between
particles in the following sense. The particles interact through
elastic collisions which, for the case of spherical particles of equal
size and mass, correspond to an exchange of the radial component of
the velocities of the two colliding particles upon impact, namely when
$|\vec{x}_i-\vec{x}_j| = 2\,a$.
 
The clustering phenomenon occurs naturally, independently of the
carrier flow compressibility, for an ensemble of particles which
evolve according to Eq.~(\ref{eq:1}), even in the absence of
interactions or collisions between particles.  We note that the case
of no collisions is equivalent to the ghost particle approach.  The
physical mechanisms which lead to strong inhomogeneities in the
spatial distribution of particles arise from the underlying
dissipative chaotic dynamics: The system is characterised by a
constant contraction rate $d/\tau$ in the position-velocity phase
space which drives the particles towards a dynamically evolving
fractal set~%\cite{fractal}%.
The clusters of particles are the projection of such fractals on the
position space.  The fractal dimensions of particles distribution
provide a convenient tool to quantify their clustering
\citep[see][]{prl_nostro,calza}.  In particular the correlation
dimension $D_2$ is defined via the power-law behaviour of the
probability distribution function (PDF) of inter-particle distances
$p_2(r) \sim r^{D_2 - 1}$ (see the inset of
Fig.~\ref{fig:pdf_no_hydro}a for particles with different Stokes
numbers in a three-dimensional random flow). Note that the density
$p_2(r)$ is such that the probability that two particles are at a
distance between $r$ and $r+\mathrm{d}r$ is $p_2(r)\,\mathrm{d}r$ and
relates to the radial distribution function $g(r)$: in three
dimensions, one has $g(r) = 4\pi\,r^2\,p_2(r)$.

The intensity of clustering is influenced by the properties of the
velocity field $\vec{u}(\vec{x},t)$ and in particular by its spatial
and temporal correlations.  Here we will assume that the velocity is
differentiable in space and time and characterised by unique time and
length scales. Let us denote by $L_\mathrm{f}$ the fluid flow
correlation length, which is assumed much larger than the particle
radius $a$, and by $\tau_\mathrm{f}$ its correlation time, which is of
the same order as the turnover time $L_\mathrm{f}/U$ where $U$ is the
typical amplitude of $\vec{u}$.  In the limit of vanishing dissipation
$St=\tau/\tau_\mathrm{f} \to 0$, Eq.~(\ref{eq:1}) becomes that of
tracers, namely $\dot{\vec{x}}_i = \vec{u}(\vec{x}_i,t)$, and the
incompressibility condition ${\bf \nabla} \cdot \vec{u}=0$ ensures a
uniform distribution of particles.  Particles distribute uniformly
also in the opposite limit $St \to \infty$, in which the force acting
on particles become very small so that they follow almost a ballistic
motion and fill the whole position-velocity phase-space.  The maximum
of clustering is achieved for intermediate (order-unity) values of
$St$ where the minimum of the fractal dimension is reached.

Clearly, the presence of collisions will affect the particle
distribution on scales of the order of their size. Heuristically, one
expects the two-particle distribution $p_2(r)$ to be unchanged at
separations $r$ much larger than the interaction distance $2\,a$.  At
separations of the order of $2\,a$, the elastic collisions decorrelate
the particle dynamics from the carrier flow. Naively, one would then
expect that the particles distribute like an ideal gas at such scales,
and consequently have a uniform distribution. In dilute systems, the
crossover between these two regimes would occur at a scale given by
the distance travelled by the particles before it relaxes to its
attractor. This distance can be written as $r_\star =
\tau\,v_\mathrm{c}(St)$, where $v_c$ is the typical velocity
difference at collisions. However, as seen from the inset of
Fig.~\ref{fig:pdf_no_hydro}a, this naive picture seems wrong. As we
will see in the next section, we indeed find that the two-particle
density $p_2$ diverges when $r\to 2\,a$. This effect is due to the
presence of sticky elastic collisions.

\section{Particle adhesion through recurrent collisions}
\label{sec:sticky}
To investigate the effect of collisions on clustering we resort to
numerical simulations of Eq.~ (\ref{eq:1}).  For simplicity and
without any loss of generality, we consider two ($N = 2$) particles in
a three-dimensional cubic domain of size $L$ with periodic boundary
conditions. The velocity field is obtained from a superposition of
Fourier modes whose amplitudes are stochastic Ornstein-Uhlenbeck
processes with Gaussian statistics and a correlation time
$\tau_\mathrm{f}$.  The amplitude of each mode, which has a standard
deviation of the order of $L/\tau_\mathrm{f}$, is chosen to ensure
statistical isotropy at small scales. In our simulations we use
several values of $\tau$ with Stokes numbers $St =
\tau/\tau_\mathrm{f}$ lying between 0.01 and 1.1. We have done
simulations with various values of the particle radius $a$, but report
here results obtained for $a \approx 0.02$. Our time marching is an
implicit Euler scheme with a fixed time step $\delta t = 10^{-4}$ when
the particles are far away from each other. However, when the particle
are close to each other, the time step is adapted in order to resolve
collisions with a high accuracy.
%%%%%%%%%%%%%%%%%%%%%%%%%%%%%%%%%%%%%%%%%%%%%%%%%%%%%%%%%%%%%%%%%%%
\begin{figure}
\begin{center}
\includegraphics[width=0.495\textwidth]{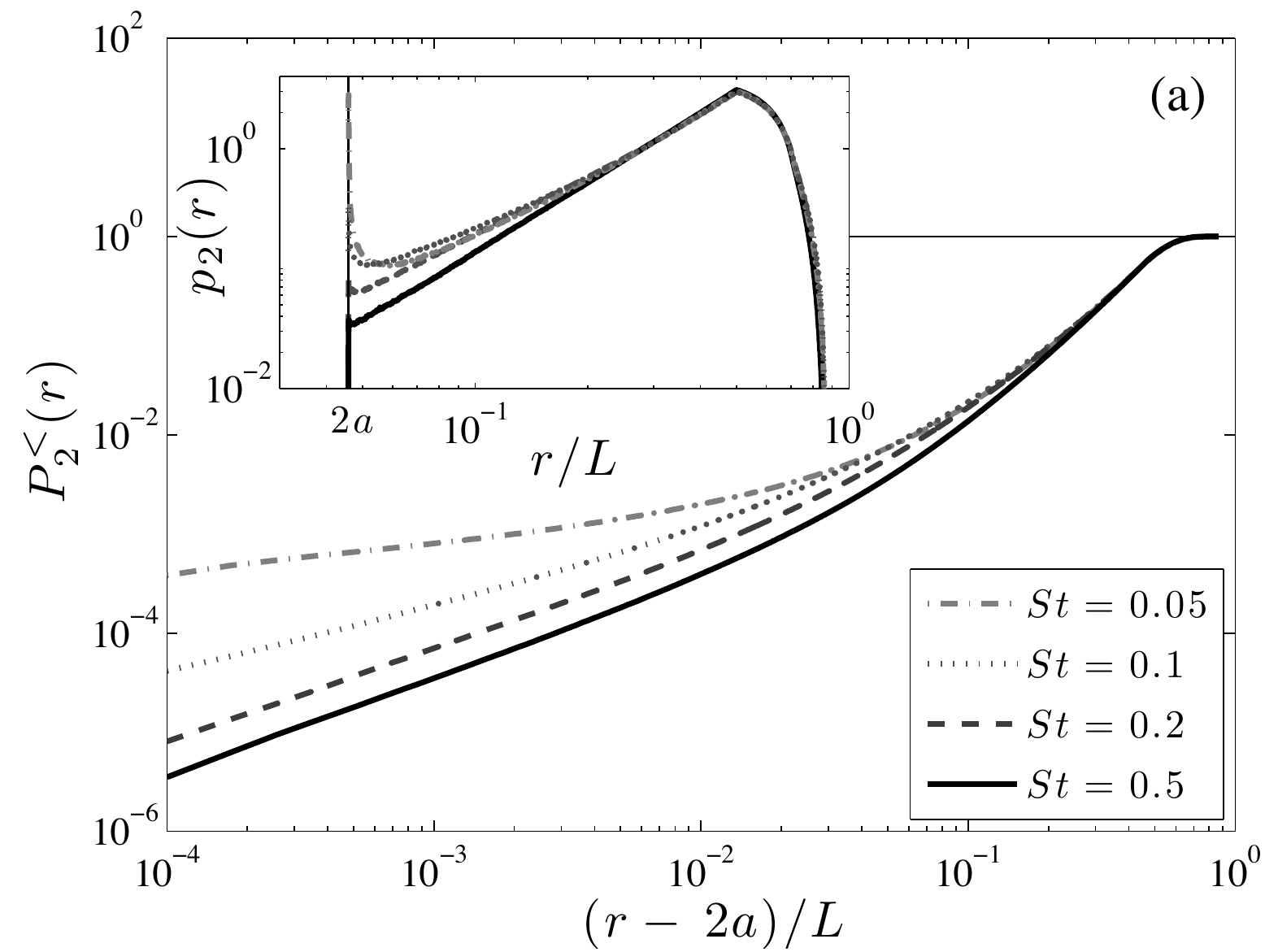}
\includegraphics[width=0.495\textwidth]{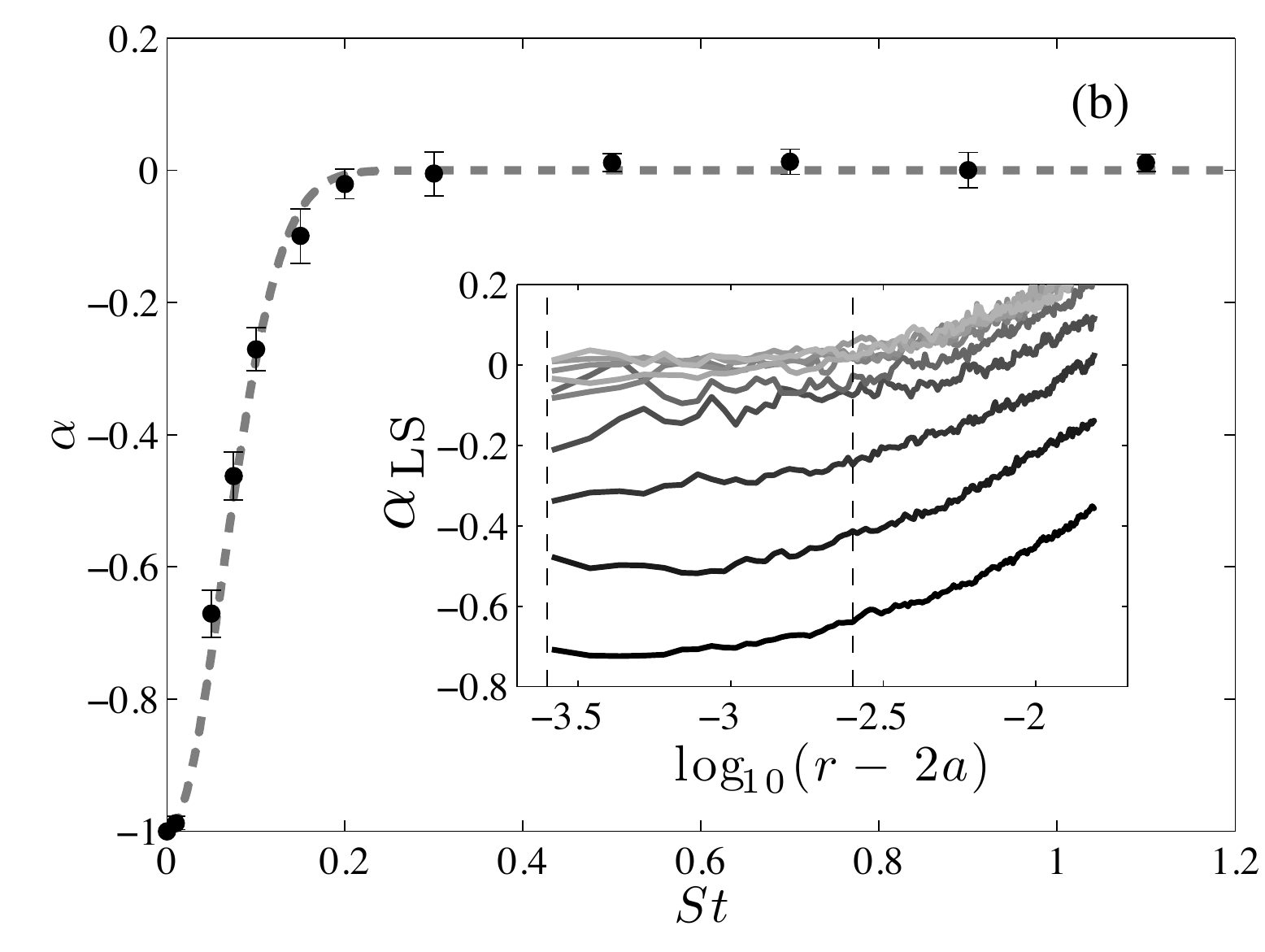}
\caption{(a) Log-log plot of the cumulative probability distribution
  $P^<_2(r)$ as a function of $(r-2a)$ for some representative values
  of $St$ as indicated in the legend. Inset: Log-log plot of the
  two-particle density $p_2(r) = \ud P^<_2(r) / \ud r$ as a function
  of $r$ for the same values of $St$. (b) Exponent $\alpha$ versus
  $St$. The gray dashed line is an empirical fit of the form
  $-\exp(-(St/0.09)^2)$. Inset: local slopes $\alpha_{\rm LS} = d\log
  P^<_2(r-2a)/d\log (r-2a) - 1$ as a function of $(r - 2a)$ for
  various $St$. The two dashed vertical lines indicate the region over
  which we calculate the mean and the standard deviation of
  $\alpha_{\rm LS}$. }
\label{fig:pdf_no_hydro}  
\end{center}
\end{figure} 
%%%%%%%%%%%%%%%%%%%%%%%%%%%%%%%%%%%%%%%%%%%%%%%%%%%%%%%%%%%%%%%%%%%

Our numerical simulations reveal a very interesting, hitherto unknown,
phenomenon which is clearly absent in the collisionless or
ghost-collision case.  As expected the effects of collisions on the
spatial distribution of particles, as characterised by the PDF of
inter-particle distance, are negligible for particle separations much
larger than $2\,a$. However collisions dramatically affect the
statistics of pair-separations at small scales.  In particular, the
probability distribution function of the inter-particle distance
$p_2(r)$ displays a power law behaviour $p_2(r) \sim (r-2\,a)^{\alpha}$
for distances close to the cutoff $2\,a$. The exponent $\alpha$ is a
monotonically increasing function in the Stokes number $St$ : It
begins with the value $-1$ for $St\to 0$ and approaches 0 as $St \to
\infty$.  Figure~\ref{fig:pdf_no_hydro}a shows the cumulative
probability distribution $P^<_2(r)$, which is the probability of
finding two particles at a distance less than $r$, as a function of
$(r-2a)$, on a log-log scale, for some representative values of $St$
that we use in our simulations. We clearly find that $P_2^<$ behaves
as a power law $\propto (r-2\,a)^{\alpha+1}$ at small values of
$r-2\,a$. From a local slope analysis we extract the local scaling
exponent $\alpha_{\rm LS} = d\log P_2^<(r-2a)/d\log (r-2a) - 1$ (see
the inset of Fig.~\ref{fig:pdf_no_hydro}b); the mean of this gives us
a measure of the scaling exponent $\alpha$ and the standard deviation
an estimate of the error.  In Fig.~\ref{fig:pdf_no_hydro}b we show the
behaviour of $\alpha$ as a function of the Stokes number. The gray
dashed line is a fit to our numerical data and is given by
$-\exp(-(St/0.09)^2)$. In the inset the dashed vertical lines denote
the region over which we calculate the mean and the standard deviation
to obtain $\alpha$. Two asymptotic values are clearly visible on
Fig.~\ref{fig:pdf_no_hydro}b: one observes that $\alpha \to 0$ when
$St \to \infty$ and $\alpha \to -1$ when $St \to 0$. Note that the
stiffness of the system in the limit $St \to 0$ prevents us from
obtaining accurate numerical results in this limit. Nevertheless, our
data show a monotonic convergence towards $\alpha = -1$.  These two
asymptotic values of the exponent are signatures of two different
collision mechanisms.

For large Stokes numbers the motion of particles is weakly correlated
with the local values of the fluid velocity field, because it is
determined by the cumulative contributions of the flow integrated over
the particle trajectories with a long memory kernel.  Close to the
collision distance individual particle velocities are almost
uncorrelated and vary on very large timescales. The relative motion of
the two particles is therefore almost ballistic, with a random
relative velocity.  The time that the two particles spend at a
distance between $r$ and $r+\mathrm{d}r$ is given by
$\mathrm{d}t=V\,\mathrm{d}r$, where $V$ is a their typical velocity
difference. This leads to $p_2(r) \sim (r - 2\,a)^0$ thus yielding
$\alpha\to 0$ when $St\to\infty$.

The limit $St \to 0$ is more complicated.  In this case the motion of
the particles is strongly correlated with the fluid velocity field
$\vec{u}$, and therefore particles typically arrive at colliding
distances with a very small relative velocity, of the order of the
fluid velocity difference, namely $2\,a\,\sigma$, where $\sigma$ is
the local gradient of $\vec{u}$. After the collision they tend to
separate but the fluid velocity field quickly brings them back
together, because of the short relaxation time $\tau \ll
\tau_\mathrm{f}$.  Particles then collide again and this mechanism
will, therefore, lead to a long series of high-frequency collisions
during which the particle remain within a distance of the order of
$2\,a$. Next, to quantify the effects of these events on the
statistics of inter-particle distance, let us consider a simple, one
dimensional model for the separation $r= |\vec{x}_1 -\vec{x}_2|$ and
the radial relative velocity $v=(\vec{v}_1 -\vec{v}_2) \cdot
{\vec{\hat{\bm r}}}$ of two particles, between two collisions occurring at
time $t_n$ and $t_{n+1}$.  The equations of motion can be written as
\begin{equation} 
\dot{r} = v \;,\;\;\; \tau \dot{v} = -v - 2\,a\,\sigma \;.
\label{eq:2}
\end{equation}
Here, we have assumed that the fluid velocity gradient $-\sigma<0$
remains constant. This is justified because particle dynamics is much
faster than the correlation time of the fluid velocity ($\tau \ll
\tau_f$). Also, we have neglected higher-order terms in the Taylor
expansion of~$\vec{u}$. The above equation gives $r(t) = 2a -\tau
(v_n+ 2 a \sigma) [\exp(-t/\tau)-1]- 2 a \sigma t$ and $v(t) =
v_n\exp(-t/\tau) + 2 a \sigma [\exp(-t/\tau)-1]$ where $v_n$ is the
relative radial velocity immediately after the $n$-th impact at time
$t_n$.  Introducing the small parameter $\epsilon_n = v_n
/(2\,a\,\sigma)$ and Taylor expanding the above expressions for
$t/\tau \ll 1$ one obtains a recursive relation for the relative
velocity at collision, namely $\epsilon_{n+1} = \epsilon_n (1-2
\epsilon_n/3 )$, which leads to $v_n \sim 2\,a\,\sigma /n$.  The
inter-collision time $\theta_n = t_{n+1}-t_n$ decreases as $\theta_n
\sim \tau/n$, and the maximum distance $r^*$ reached by the two
particles in the excursion between two collisions scales as $\delta =
r^*_n/(2a)-1 \sim \sigma \tau /n^2$.  The number of collisions
increases exponentially in time $n_c \sim \exp(C\,t/\tau)$, with
$C>0$, and the inter-particles distance goes to 0.  We call this
phenomenon {\it sticky elastic collisions}. Note that the series
defined by the sum of inter-collision times $\sum_n\theta_n$ is not
converging. This ensures that the number of collisions does not become
infinite in a finite time, at variance with the case for inelastic
collisions where the particles eventually collapse and aggregate. In
the case of the one-dimensional sticky elastic collisions, the
recurrent process stops only when the fluid velocity gradient becomes
positive (for $t \simeq \tau_\mathrm{f}$) and takes the two particles
far away.  During any one of these events, only the first $m$
collisions will contribute to the probability of having the two
particles at a distance larger than $r\simeq 2 a (1+ \sigma \tau
/m^2)$.  The fraction of time spent at a distance larger that $r$ is
thus $\sim \sum_{n<m} \theta_n \sim \ln m \sim -\ln (r-2a)$. This gives
for the PDF of inter-particle distances $p_2(r) \sim (r-2\,a)^{-1}$,
yielding the asymptotic value $\alpha \to-1$ for $St\to0$.

The heuristic arguments developed here to quantify the statistical
signature of sticky events are purely one-dimensional. However, they
extend to higher dimensions by geometrical considerations. The
one-dimensional case would hold true if the velocity difference
$\vec{v}$ between the particles were exactly aligned with their
separation $\vec{r}$. A misalignment leads to rebounds of the two
particles at different locations on their surfaces. For spherical
particles, this implies that better is the alignment between $\vec{v}$
and $\vec{r}$, higher is the number of successive secondary
collisions. Statistically, this implies that the distribution of
distances is dominated by almost head-on collisions, which can
essentially be treated as a one-dimensional problem. The space
dimensionality should just appear as a multiplicative factor in the
power-law behaviour at $r\to 2\,a$.

%%%%%%%%%%%%%%%%%%%%%%%%%%%%%%%%%%%%%%%%%%%%%%%%%%%%%%%%%%%%%%%%%%%
\begin{figure}
\begin{center}
\includegraphics[width=0.495\textwidth]{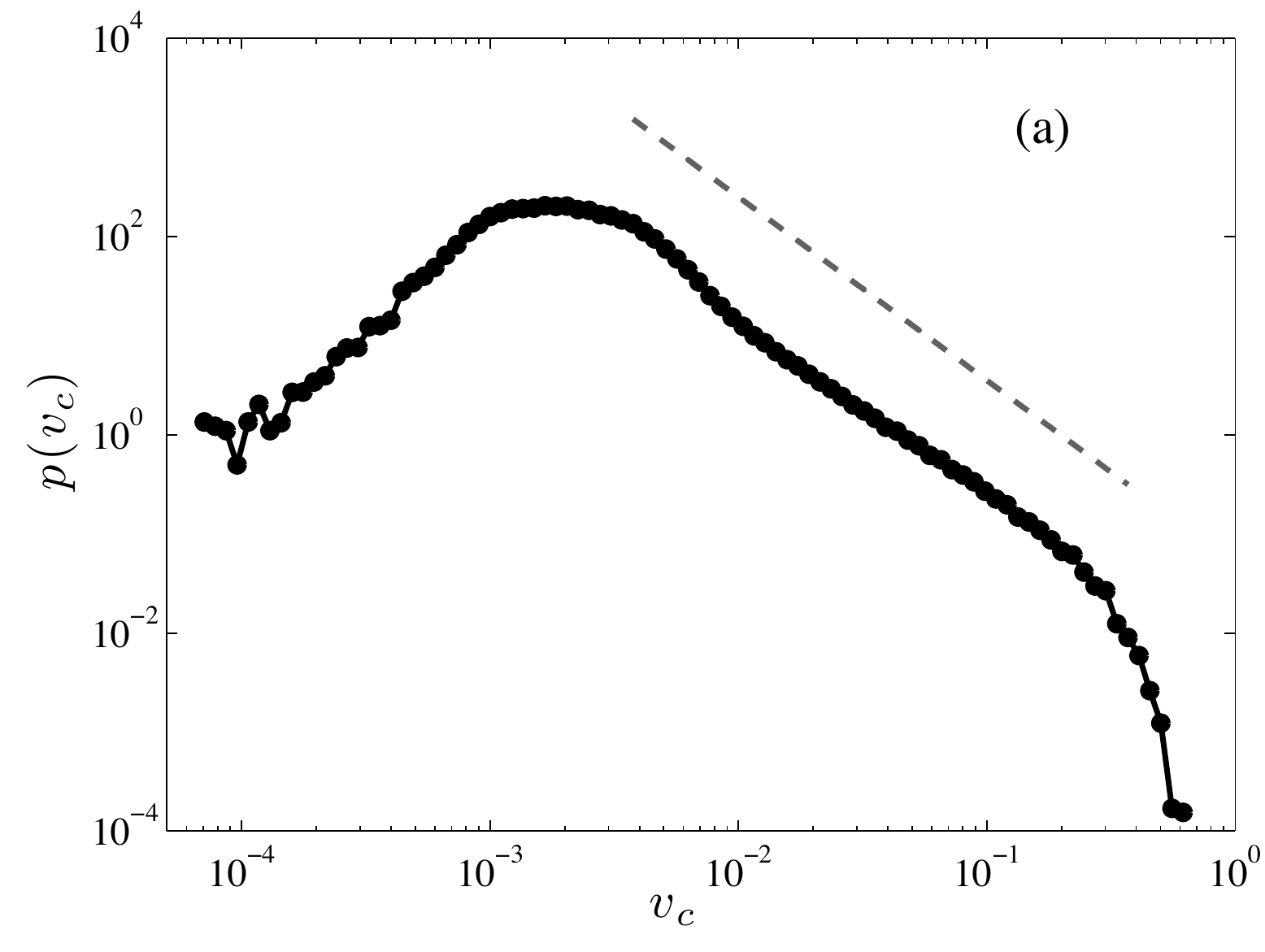}
\includegraphics[width=0.495\textwidth]{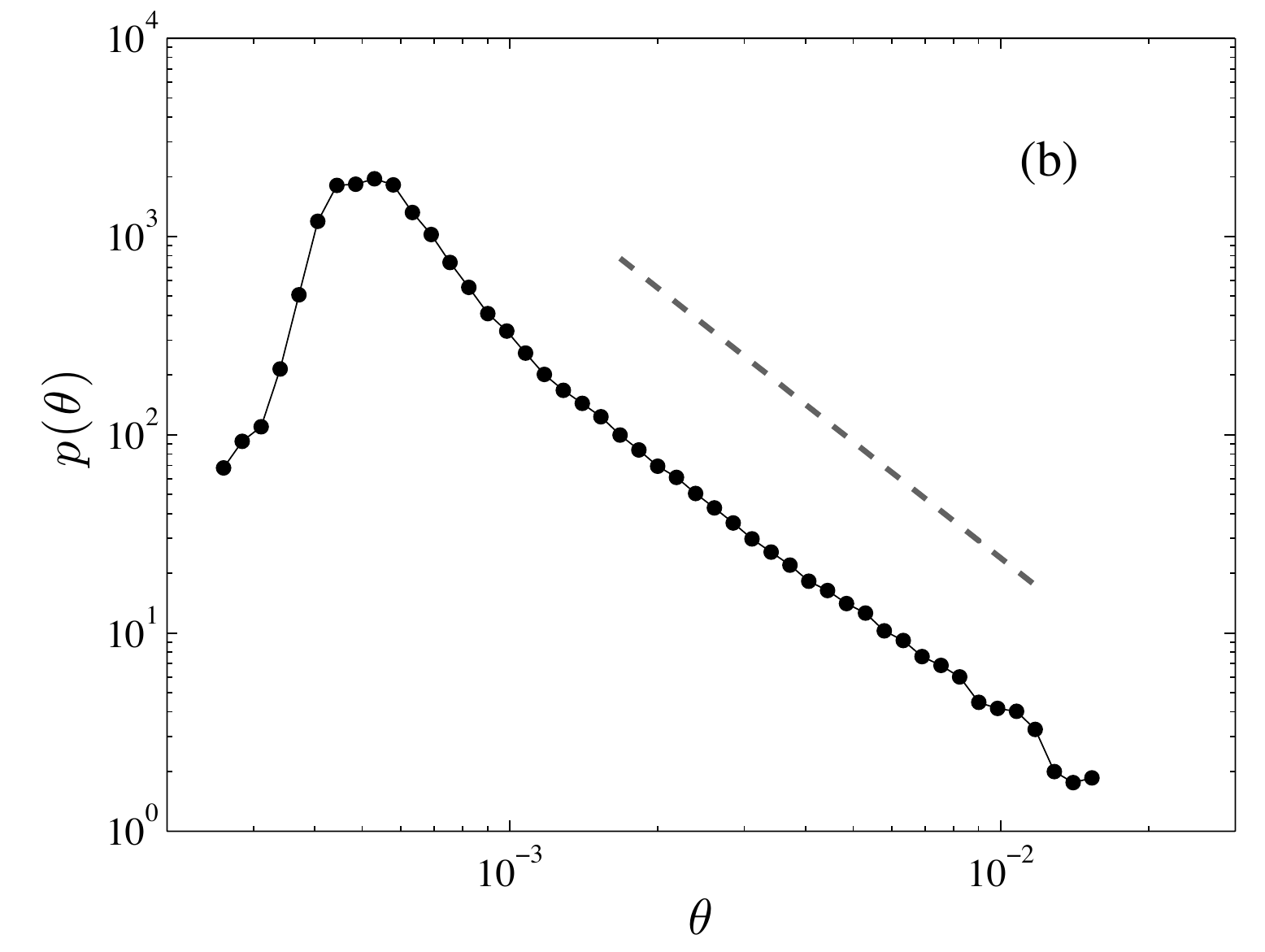}
\caption{(a) Log-log plot of the PDF of inter-collision radial
  velocity $v_c$ for $St = 0.01$; the dashed line shows a scaling of
  $-1.85$. (b) Log-log plot of the PDF of inter-collision time
  $\theta$ for $St = 0.01$; the dashed line shows a scaling of
  $-1.95$.}
\label{fig:inter_coll_stats}  
\end{center}
\end{figure} 
%%%%%%%%%%%%%%%%%%%%%%%%%%%%%%%%%%%%%%%%%%%%%%%%%%%%%%%%%%%%%%%%%%%

The effects of these events are detectable also in the statistics of
inter-collision times $\theta$ and relative radial velocity at
collision $v_c$.  Within the model derived above, for $St \to 0$, we
obtain $p(v_c) \sim v_c^{-2}$ and $p(\theta) \sim \theta^{-2}$.  In
Fig.(\ref{fig:inter_coll_stats}a) and
Fig.(\ref{fig:inter_coll_stats}b) we show a log-log plots of $p(v_c)$
versus $v_c$ and $p(\theta)$ versus $\theta$, respectively, obtained
from simulations for $St = 0.01$; the dashed lines shows a scaling
exponent of $\approx -1.9$ in each case. We have checked the scaling
becomes significantly shallower and the extent of scaling gets
progressively reduced with $St$ increases. Note that a similarly steep
increase of the inter-collision time distribution at small values has
been observed by \cite{ten2004fully} in full direct numerical
simulations of finite-size particles suspended in a viscous flow. This
effect has been interpreted as a possible consequence of lubrication
forces between the particles that makes the particles remain close to
each other for long times. In the next section, we will comment on the
influence of hydrodynamical interactions on sticky elastic
collisions. Nevertheless it is worth stressing here that lubrication
is not necessary to obtain multiple collisions between particles.

The tails of $p(v_c)$ and $p(\theta)$ at small values seem to indicate
that they cannot be normalised. However, we observe for small but
finite values of $St$, a cutoff at the smallest values of $v_c$ and
$\theta$, which prevents this divergence.  Indeed, the power-law
behaviours are due to typical sticky events. As we have seen, the
number of collisions is of the order of $n_c
\sim\exp(C/St)$. Therefore, the minimal collisional velocity and
inter-collision time are both $\propto 1/n_c\sim\exp(-C/St)$. The
power-law is thus just appearing in intermediate ranges, namely
$(2\,a/\tau_\mathrm{f}) \exp(-C/St) \ll v_c \ll
(2\,a/\tau_\mathrm{f})$ and $\tau \exp(-C/St) \ll \theta \ll
\tau$. The events leading to values smaller than the lower bounds are
related to situations where the fluid velocity gradient is maintained
negative for an exceptionally long time. This happens with an
inverse-Gaussian probability in a finite-correlation-time flow, whence
the cutoff. Also, the extension of this argument to the singular limit
$St=0$ is far from obvious. Note finally that in dimensions higher
than one, the geometrical considerations explained above should also
provide a cut-off for very small collision times and velocities.

\section{Effect of hydrodynamic interactions}
\label{sec:hydro}
In this section, we address the question whether or not sticky elastic
collisions can be observed in realistic flows. For that, it is
important to know how far the observations and conclusions drawn above
are valid when we, in addition to collisions, introduce hydrodynamic
interactions between particles. To answer this question, we take into
account the far--field or long--range interactions and assume that it
is valid all the way to the smallest separations.  This approach
yields several interesting results which we discuss below. Of course
this assumption, in reality, breaks down when particles approach {\it
  very} close to one other. However in that case, either lubrication
leads to an effective increase in the particle physical radius or, in
the case when the velocity difference is too large, hydrodynamics
might not be a valid description of the interactions between the
particles.

In its simplest formulation, the long--range hydrodynamic interactions
between particles in a flow is taken into account by considering the
perturbation in the ambient fluid velocity field, as experienced by an
individual particle, because of the motion of all the other particles.
Thus the effective velocity field acting on any particle is a
superposition of the unperturbed (turbulent) advecting flow
$\vec{u}(\vec{x}, t)$ and of the perturbation to this flow due to the
other particles. Formally, this perturbation $\vec{u}^{(i)}$ on the
{\it isolated} $i$th particle due to another particle $j$ of radius
$a$ at a distance $|\vec{r}^{(j)}|$ and moving with a velocity $\vec{v}^{(j)}$
can be written as a combination of a Stokeslet and a potential dipole
flow as follows \citep[see][]{Wang}:
\begin{equation}
  \vec{u}^{(i)} = \left [ \frac{3}{4} \frac{a}{r^{(j)}} -
    \frac{3}{4}\frac{a^3}{|\vec{r}^{(j)}|^3}\right
  ]\frac{\vec{r}^{(j)}}{|\vec{r}^{(j)}|^2}\left
    (\vec{v}^{(j)}\cdot\vec{r}^{(j)}\right ) +  \left [ \frac{3}{4}
    \frac{a}{|\vec{r}^{(j)}|} +
    \frac{1}{4}\frac{a^3}{|\vec{r}^{(j)}|^3}\right ] \vec{v}^{(j)}.
  \label{eq:fullhydro}
\end{equation}
In a system of $N$ particles, the net perturbation on the flow field
experienced by any particle $i$ is obtained by summing over the
contributions made by each of the other $(N -1)$ particles. Given the
structure of the equations, for a system of $N$ particles, it is
impossible to solve exactly the perturbation field. Thus various
approximations and iterative schemes become essential. However, in the
present problem being studied in this paper, which involves two
particles only, it is possible to solve exactly the hydrodynamic
interaction term as it involves merely an inversion of a 6 $\times$ 6
matrix.

Let us now try to understand the effect of hydrodynamic interactions
on the particle dynamics from a theoretical point of view. Without any
loss of generality, and considering only two-particle interactions,
let us consider a model where the two particles approach each other
with the same velocity $\vec{v}$. A direct consequence of this model
is that the perturbation $\vec{u}^{(1)}$ on particle 1 due to particle
2 is equal and opposite to the perturbation $\vec{u}^{(2)}$ on
particle 2 due to particle 1, i.e.\ $\vec{u}^{(1)} =
-\vec{u}^{(2)}$. The equation of motion for particle~1 can be written
as
\begin{eqnarray}
  \frac{\mathrm{d}\vec{v}}{\mathrm{d}t} &=& -\frac{1}{\tau}\left
    (\vec{v} - \vec{u}_1 - \vec{u}^{(2)} \right ); \\
  \label{hydro-dvdt}
\vec{u}^{(2)} &=& \left ( \vec{v} + \frac{1}{2}\,\vec{\sigma}\,\vec{r}  -
  \vec{u}^{(1)} \right )\left (\frac{3}{2}\frac{a}{|\vec{r}|} -
  \frac{1}{2}\frac{a^3}{|\vec{r}|^3}\right ),
\end{eqnarray}
where $\vec{\sigma}$ denotes the unperturbed fluid velocity gradient.
Since, at the point of collision, $|\vec{r}| = 2a$, one has
\begin{equation}
  \vec{u}^{(1)} = -\frac{11}{5}\left ( \vec{v} - 2\,
    \vec{\sigma}\,{\bm\hat{\bm r}}\,a \right ).
  \label{hydro}
\end{equation}
By using Eq. (\ref{hydro}) in Eq. (\ref{hydro-dvdt}), we eventually obtain 
\begin{equation}
  \frac{\mathrm{d}\vec{v}}{\mathrm{d}t} = -\frac{16}{5\tau}\left (
    \vec{v} - 2\,\bm{\sigma}\,\bm\hat{\bm r}\,a \right ).
  \label{final-hydro-dvdt}
\end{equation}
The above analysis shows that the effect of long-range hydrodynamic
interactions reduces in the vicinity of collisions to the dynamics in
absence of interactions but with an effective Stokes number, which is
equal to the actual Stokes number reduced by a factor of $16/5 =
3.2$. Thus a system of particles with a Stokes number $St$ and
subject to hydrodynamic interactions can be replaced by a system of
particles, {\it without} any hydrodynamic interactions but with an
effective Stokes number $St_{\rm eff} = (5/16)\,St$ when we consider
their statistical properties for very small inter-particle
separations.

%%%%%%%%%%%%%%%%%%%%%%%%%%%%%%%%%%%%%%%%%%%%%%%%%%%%%%%%%%%%%%%%%%%
\begin{figure}
\begin{center}
\includegraphics[width=0.495\textwidth]{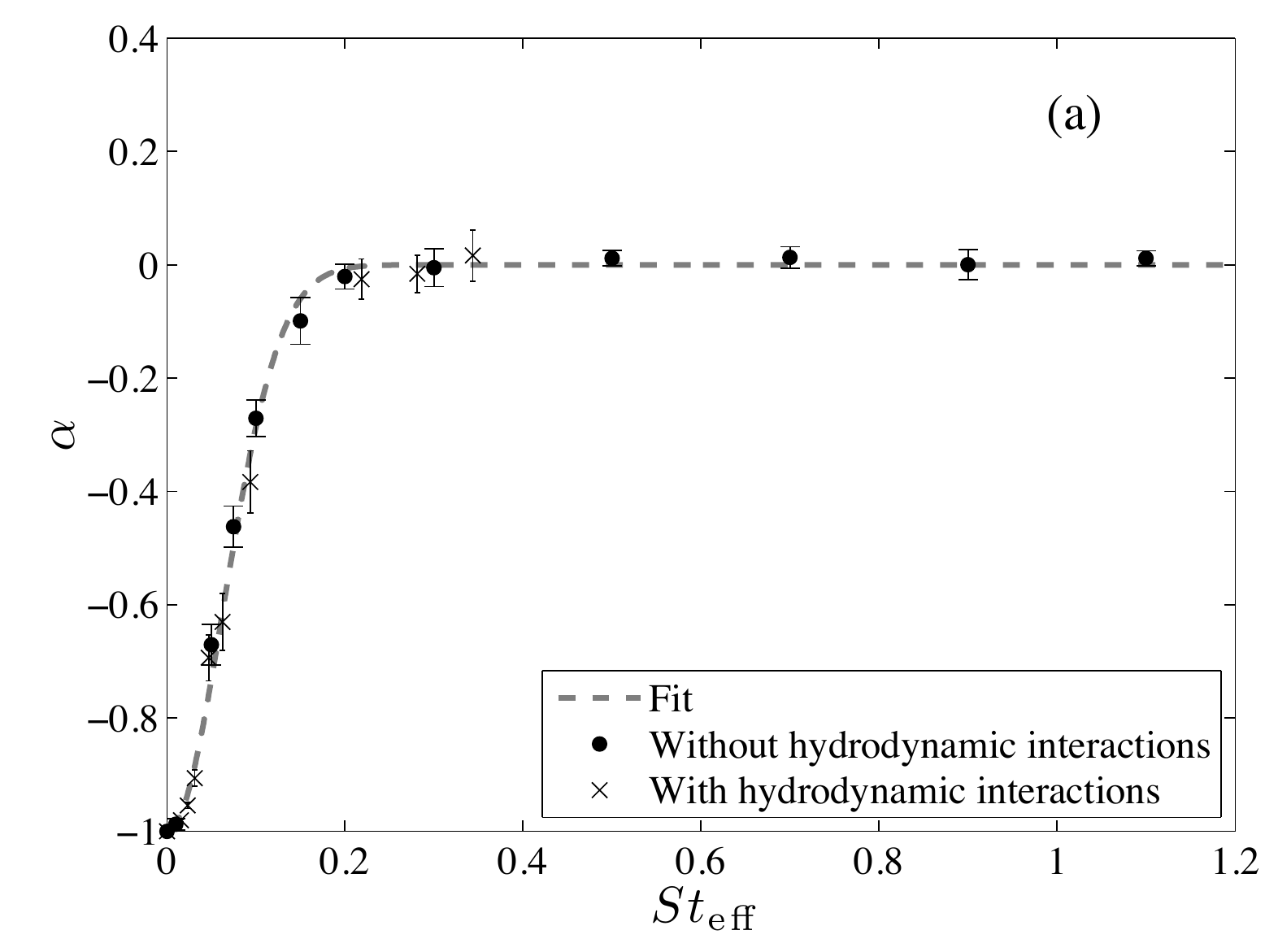}
\includegraphics[width=0.495\textwidth]{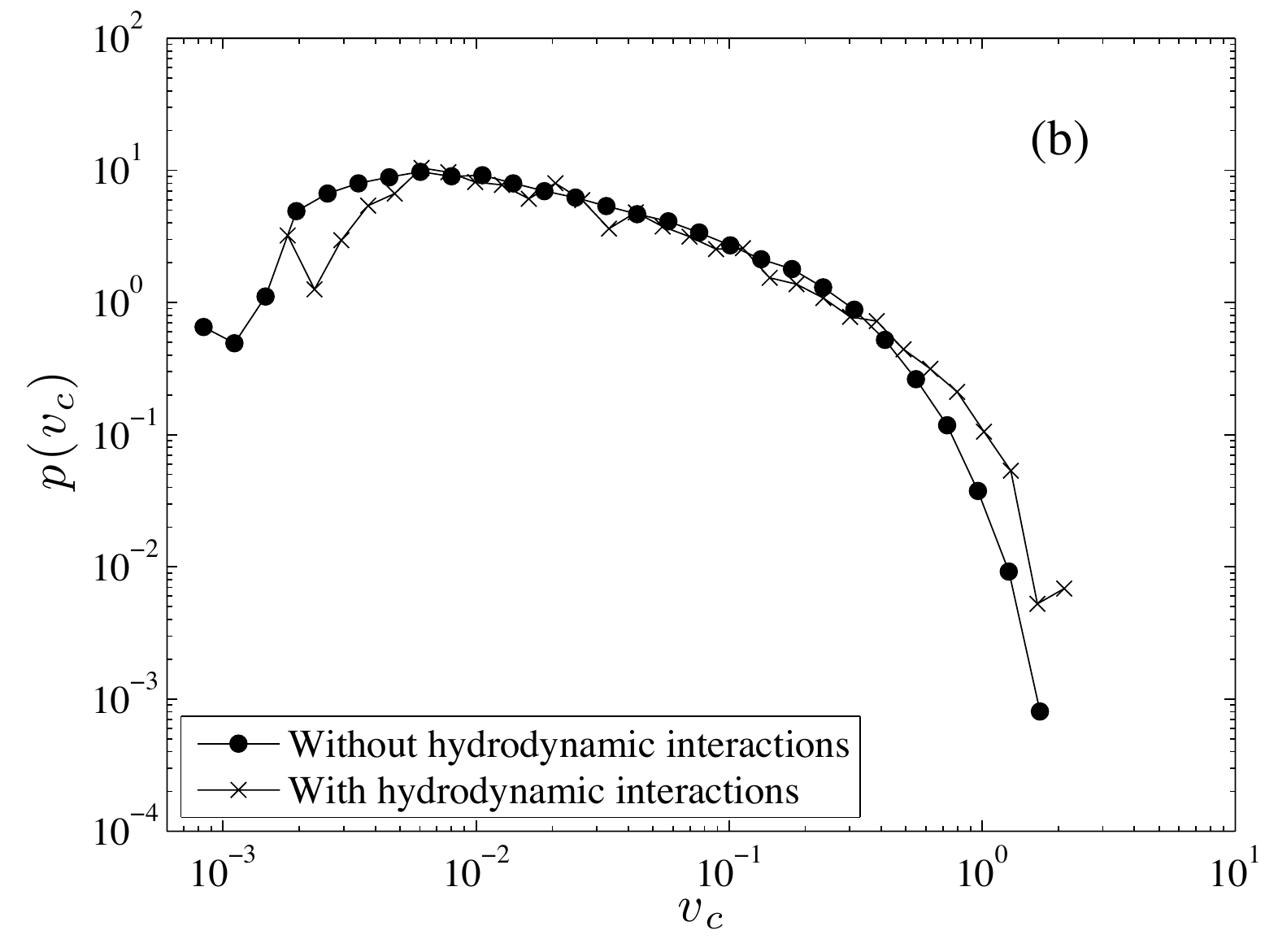}
\caption{(a) Exponents $\alpha$ of the two-particle distribution
  $p_2(r) \propto (r-2\,a)^\alpha$, along with their error bars. The
  case {\it with} hydrodynamic interactions ($\times$) is represented
  versus $St_{\rm eff} = (5/16)\,St$ and that {\it without}
  interactions ($\bullet$), as a function of $St$; the gray dashed
  line is the empirical fit $-\exp(-(St/0.09)^2)$ discussed in
  previous section. (b) PDF of inter-collision radial velocity $v_c$
  obtained from simulations with hydrodynamic interactions ($\times$)
  for $St = 0.15$, that is $St_{\rm eff} = 0.047$ and without
  hydrodynamic interactions ($\bullet$) for $St = 0.05$. }
\label{fig:hydro}  
\end{center}
\end{figure} 
%%%%%%%%%%%%%%%%%%%%%%%%%%%%%%%%%%%%%%%%%%%%%%%%%%%%%%%%%%%%%%%%%%%

To confirm the arguments presented above we resort once more to
numerical simulations by implementing the long-range hydrodynamic
interactions \eqref{eq:fullhydro}. We use values of $St$ between 0.05
and 1.1 as we had used for the case without any interaction terms. We
begin by measuring the values of the exponent $\alpha$ introduced in
the previous Section that describes the behaviour of the
inter-particle distance distribution.  Figure~(\ref{fig:hydro}a) shows
the values of $\alpha$ obtained as a function of the effective Stokes
number $St_{\rm eff} = (5/16)\,St$ (crosses); our data seems to fall,
within error bars, on the empirical fit shown by a dashed gray
line. To make the comparison more illuminating we plot on the same
graph and as a function of the actual Stokes number $St$, the values
of $\alpha$ (black dots) obtained from numerical simulations without
hydrodynamic interactions, and already shown in
Fig.~(\ref{fig:pdf_no_hydro}b). Furthermore, in
Fig.~(\ref{fig:hydro}b) we show the PDF of the inter-collision
velocity $p(v_c)$ as a function of $v_c$, on a log-log scale, for $St
= 0.05$ obtained from a simulation {\em without} hydrodynamic
interactions (black dots) and for $St = 0.15$ obtained from a
simulation {\em with} hydrodynamic interactions (crosses). The two
PDFs are nearly overlapping as our arguments before would suggest that
the effective Stokes is $St_{\rm eff} = 0.047 \approx 0.05$ for the
case with the hydrodynamic interactions.

Our results suggest that hydrodynamic interactions increase the
efficiency of dissipative mechanisms in terms of a reduction of the
effective Stokes number. However, such considerations can only lead to
qualitative deductions as our study account for long-range
interactions only. The effect of lubrication forces will become
dominant for particles at very small separations. On the one hand,
this type of interaction is expected to decrease the collision
efficiency between particles
\citep[see,e.g.,][]{wang2005theoretical}. On the other hand,
lubrication is expected to increase damping when particles get close to
each other; this effect is usually modelled by a restitution
coefficient less than unity. Because of these two competing
mechanisms, it is difficult predict whether short-range hydrodynamic
interactions will enhance or diminish the sticky elastic collision
phenomenon.

\section{Conclusions}
\label{sec:concl}
In our work we have considered the effect of elastic collisions on the
clustering of inertial particles. In particular we have investigated
their influence on the probability distribution of inter-particle
distance. Surprisingly, our findings differ markedly from the naive
picture that collisions might only introduce a small-scale molecular
chaos.  We observe that the small-distance statistics is dominated by
a phenomenon, which we call \emph{sticky elastic collisions}, during
which particles undergo a very large number of collisions during a
time of the order of the fluid correlation time. It is interesting to
note that these sticky elastic collisions remarkably resembles
inelastic collapses observed in granular media, even though the
underlying assumption in granular media (conservative inter-collision
dynamics and dissipative collisions) is exactly the opposite of what
we have considered here. In addition we have investigated the
effect on this phenomenon of long-range hydrodynamic interactions
between particles.  Our results seem to indicate that the most
significant effect at small scales of such interactions is to
introduce an effective Stokes number. The problem of investigating the
effect on sticky elastic collisions of short-range hydrodynamical
interactions requires more rigorous theoretical understanding and more
elaborate numerical simulations.

In this study we have focused on two-particle interactions. It is
clear that the phenomena of sticky elastic collisions will be present
even for large numbers of interacting particles. In this light,
collective phenomena may emerge and provide a new mechanism to
dissipate kinetic energy in violent collisions between
particles. There are still many open questions concerning the
stability of coalescence processes for the high impact velocities that
are observed in turbulent settings. In particular, estimates on
relative velocities between meter-sized objects in circum-stellar
disks are by far too large to allow for their accretion and growth to
form planet embryos \citep{wurm-blum-etal:2001}. The dissipative
mechanisms relating to sticky elastic collisions might play a role
there.

\begin{acknowledgments}
  We would like to thank G.\ Falkovich and D.\ Mitra for useful
  discussions and acknowledge support from the European Cooperation in
  Science and Technology (EU COST) Action MP0806. The research leading
  to these results has received funding from the European Research
  Council under the European Community's Seventh Framework Program
  (FP7/2007-2013, Grant Agreement no. 240579) and from the Agence
  Nationale de la Recherche (Programme Blanc ANR-12-BS09-011-04).
\end{acknowledgments}

\bibliographystyle{jfm.bst}

\end{document}